\documentclass[11pt, a4paper]{article}
\usepackage{}

\usepackage{indentfirst}
\usepackage{lineno}
\usepackage{graphicx}
\usepackage{ae}
\usepackage{amsmath}
\usepackage{amssymb}
\usepackage{latexsym}
\usepackage{url}
\usepackage{epsfig}
\usepackage{cite}
\usepackage{mathrsfs}
\usepackage{amsfonts}
\usepackage{amsthm}
\usepackage{float}
\usepackage{booktabs}
\usepackage{subfigure}

\usepackage{color}

\long\def\delete#1{}

\newcommand{\be}{\begin{equation}}
\newcommand{\ee}{\end{equation}}
\newcommand{\ben}{\begin{equation*}}
\newcommand{\een}{\end{equation*}}
\newcommand{\bea}{\begin{eqnarray}}
\newcommand{\eea}{\end{eqnarray}}
\newcommand{\bean}{\begin{eqnarray*}}
\newcommand{\eean}{\end{eqnarray*}}

\newtheorem{thm}{Theorem}[section]

\newtheorem{defn}[thm]{Definition}

\numberwithin{equation}{section}

\title{EMH: Extended Mixing H-index centrality for identification important users in social networks based on neighborhood diversity \thanks{Supported by the National Natural Science Foundation of China (No.11361033) and the Natural Science Foundation of Gansu Province (No.1212RJZA029).}}

\author{Pengli Lu\thanks{Corresponding author. E-mail addresses: lupengli88@163.com (\textbf{P. Lu}), dongchen199508@163.com (\textbf{C. Dong}).} \;and\; Chen Dong
\\
\footnotesize{School of Computer and Communication, Lanzhou University of Technology, Lanzhou, 730050, Gansu, P.R. China}}

\date{}

\begin{document}

\openup 0.5\jot
\maketitle
%\linenumbers

\begin{abstract}
The rapid expansion of social network provides a suitable platform for users to deliver messages. Through the social network, we can harvest resources and share messages in a very short time. The developing of social network has brought us tremendous conveniences. However, nodes that make up the network have different spreading capability, which are constrained by many factors, and the topological structure of network is the principal element. In order to calculate the importance of nodes in network more accurately, this paper defines the improved H-index centrality ($IH$) according to the diversity of neighboring nodes, then uses the cumulative centrality ($MC$) to take all neighboring nodes into consideration, and proposes the extended mixing H-index centrality ($EMH$). We evaluate the proposed method by Susceptible-Infected-Recovered ($SIR$) model and monotonicity which are used to assess accuracy and resolution of the method, respectively. Experimental results indicate that the proposed method is superior to the existing measures of identifying nodes in different networks.

\bigskip


\bigskip
\end{abstract}

%=============================================I=n=t=r=o=d=u=c=t=i=o=n===================================================================
\section{Introduction}
In the digital multimedia era, people can get information and send messages to each other through social network, it has been widely used in transportation, health care and finance. The emergence of social network has greatly facilitated people's lives, and also gradually replaces traditional methods of communication such as television, radio and newspaper \cite{1,2,3,4}. In the social network, the spreading of information is often controlled by the set of important influential nodes. How to determine these nodes has been a hot topic of research \cite{5,6,7}. In order to solve this problem, scholars studied it from various aspects and finally divided the problem into two parts: first, calculating the importance of nodes and ranking them, then finding the most significant node's set as the initial node to spread information so as to maximize the process \cite{8,9,10}.

Social networks can be abstracted into graphs, with nodes representing users and edges denoting the connection between users. The topological location of nodes in the graph and characteristic of social networks are the two main factors that determine the importance of nodes. Due to the lack of contextual information of nodes in the network, the topological location of nodes is usually the only indicator that determines its spreading capability \cite{11,12,13}. Many classical methods have been proposed to evaluate the spreading ability of nodes in complex networks. Degree centrality \cite{14}, betweeness centrality \cite{15}, closeness centrality \cite{16} and H-index centrality \cite{17} are the most widely used measures. Degree centrality is the most direct and efficient measure, but it cannot accurately reflect the performance of the neighbor nodes. Both betweenness centrality and closeness centrality are well known global measures, but due to their high computational complexity, they cannot be applied to large-scale networks. As a mixed quantization index, H-index centrality takes the degree of nodes and their neighbors into comprehensive consideration, and there is also the problem that different nodes are given the same weight. Recently, Kitsak et al. have found that the most important nodes are those at the core of the network, based on which K-core decomposition centrality is proposed \cite{18}. However, the k-shell decomposition tends to assign many nodes with an identical k-shell index, although the spreading capability of the nodes that reside in the same k-shell may differ from each other. After that, many methods were used to improve the K-core decomposition centrality and H-index centrality. Zeng et al. proposed a method to consider the incorporating the residual degree and the exhausted degree in K-core decomposition, and it was difficult to achieve the better results because the suitable value of $\lambda$ could not be found \cite{19}. Bae and Kim proposed the measure to estimate the spreading capability of nodes in the network by summing up k-shell values of all neighbors \cite{20}. Liu et al. proposed the measure to identify and rank influential nodes by taking into account of h-index values of the node itself and its neighbors in the network \cite{21}. In conclusion, it is still an open issue to propose an effective method to identify the importance of nodes.

Information can be propagated to different parts of network through node's neighbors. The number of node's neighbors has no decisive influence on the
importance of node, while the diversity of neighboring nodes has become a new breakthrough. It is an original idea to classify nodes according to their
neighboring nodes importance and topological properties \cite{10,22,23}. Here, we propose the Extended Mixing H-index centrality ($EMH$), an improved H-index centrality ($IH$) based on the diversity of node's neighbors, combined with the cumulative centrality \cite{24} ($MC$) so that all nodes can be considered, and the topological structure of network can also be fully applied. Susceptible-Infected-Recovered ($SIR$) model and monotonicity are used to evaluate
the effectiveness of the proposed method. Experimental results in a series of networks show that $EMH$ centrality can acquire a unique ranking list and
obtain the more accuracy importance of nodes than existing measures.
%==========================================P=r=e=l=i=m=i=n=a=r=i=e=s================================================================================================

\section{Related work}\label{Se2}
In this paper, an undirected social network consisting of $|V|$ vertices and $|E|$ edges can be represented by $G(V,E)$, and the relationship between nodes can be described as the adjacency matrix $A=(a_{ij})_{|V| \times |V|}$, if node $i$ is connected to node $j$ then $a_{ij}$=1, $a_{ij}=0$ otherwise. $N_{i}$ is the set of neighbors of node $i$.

Degree centrality ($DC$) \cite{14} is a measure that only consider the topology of nodes themselves. K-shell decomposition ($KS$) \cite{18} is a global evaluation method of term-by-term cutting network, which can divide nodes into diverse shells depending on their positions. On the basis of k-shell decomposition, researchers put forward a ranking method-neighborhood coreness $cn$ \cite{20} that comprehensively consider the degree and coreness of a node, which is defined as:
    \begin{equation}\label{adjmatix}
    \begin{aligned}
    \begin{split}
    cn(i)=\displaystyle \sum_{j \in N_{i}}ks(j)
    \end{split}
    \end{aligned}
    \end{equation}
where $ks(j)$ is the k-shell value of node $j$.

Weight neighborhood centrality ($c_{i}(\phi)$) \cite{25} is an evaluation measure based on benchmark centrality $\phi$, which is defined as:
    \begin{equation}\label{adjmatix}
    \begin{aligned}
    \begin{split}
    c_{i}(\phi)=\phi_{i}+\sum_{j \in N_{i}}\frac{A_{ij}}{<A>} \ast \phi_{i}
    \end{split}
    \end{aligned}
    \end{equation}
where $A_{ij}=(k_{i}k_{j})^{\alpha}$, $k_{i}$ and $k_{j}$ are the degree of node $i$ and $j$ respectively, $\alpha$ is a tunable parameter in range $(0,1)$. $<A>$ is the average value of all $A_{ij}$ in the network. In this paper, two classical methods, degree centrality and k-shell decomposition were selected as the benchmark centrality and denoted as $cdc$ and $cks$, respectively.

Influenced by Newton's classical gravitation formula, the k-shell value of nodes can be regarded as mass, and the shortest distance between two nodes in the network can be viewed as their distance \cite{26}. Newton's classical gravitation formula is defined as:
    \begin{equation}\label{adjmatix}
    \begin{aligned}
    \begin{split}
    G(i)=\displaystyle \sum_{j \in M_{i}}\frac{ks(i)ks(j)}{d_{ij}^2}
    \end{split}
    \end{aligned}
    \end{equation}
where $M_{i}$ represents the node's neighboring set of node $i$, which the shortest path length less than or equal to the given length (i.e., $d_{ij} \leq r$, considering the scale of network, we set $r=3$), and node $j$ is an element of $M_{i}$. $ks(i)$ and $ks(j)$ are the k-shell value of node $i$ and $j$ respectively. $d_{ij}$ is the shortest distance between two nodes $v_{i}$ and $v_{j}$. On this basis, researchers put forward the improved Newton's gravity centrality \cite{27}: the k-shell value of node itself is considered as its mass, and the degree of neighboring nodes is taken into account as its mass, the improved Newton's gravity centrality can be defined as:
    \begin{equation}\label{adjmatix}
    \begin{aligned}
    \begin{split}
    IGC(i)=\displaystyle \sum_{j \in M_{i}}\frac{ks(i)k(j)}{d_{ij}^2}
    \end{split}
    \end{aligned}
    \end{equation}
where $k_{j}$ represents the degree of node $j$.

Due to k-shell decomposition method and degree centrality have different effect in complete and incomplete global network, weighted k-shell degree neighborhood centrality had been put forward \cite{28}. The method is consisted of two parts: in the first part, the k-shell value and degree of the node are used to estimate the edge weight. In the second part, calculates the value of each node in the network. The two parts are defined as follows:
    \begin{equation}\label{adjmatix}
    \begin{aligned}
    \begin{split}
    w_{ij}=\{ (\alpha \ast d_{i} + \mu \ast core_{i}) \ast (\alpha \ast d_{j} + \mu \ast core_{j}) \}
    \end{split}
    \end{aligned}
    \end{equation}
where $j$ is a member of the set of neighbors of node $i$. $core_{i}$ and $core_{j}$ are the k-shell value of node $i$ and $j$, respectively. $d_{i}$ and $d_{j}$ are the degree of node $i$ and $j$, respectively. $\alpha$ and $\mu$ are the tunable parameters in range (0,1).
    \begin{equation}\label{adjmatix}
    \begin{aligned}
    \begin{split}
    ksd_{i}^w=\displaystyle \sum_{j \in N_{i}}w_{ij}
    \end{split}
    \end{aligned}
    \end{equation}
where $w_{ij}$ is the edge's weight between the nodes $i$ and $j$ .
%=================================B=a=s=i=c====M=a=t=r=i=x========================================================================================================================

\section{Proposed method}\label{Se3}
\begin{defn}
{\em
The neighbor diversity of node $v$, referring to the degree and H-index of the node, is defined as:
}
\end{defn}
    \begin{equation}\label{adjmatix}
    \begin{aligned}
    \begin{split}
    neighbor-diversity(v)={\displaystyle \sum_{j=1}^{H_{max}} \alpha_{j}(v)}
    \end{split}
    \end{aligned}
    \end{equation}
where $\alpha_{j}(v)$ is $1$ if a member of the set of neighbors of node $v$ is in the $jth$ H-value, and is $0$ otherwise.

In social network, the distribution of nodes is uneven, and the number of neighboring nodes cannot accurately reflect the importance of nodes. The dispersion degree of nodes also exposes the same problem. Due to the low diversity and distribution of neighboring nodes in the network, the influence diffusion range of nodes is also limited.

\begin{defn}
{\em
The Improved H-index centrality of node $v$ ($IH(v)$), referring to the degree and neighbor-diversity of the node, is defined as:
}
\end{defn}
    \begin{equation}\label{adjmatix}
    \begin{aligned}
    \begin{split}
    IH(v)=\frac{(\alpha_{1} \ast A_{1})+(\alpha_{2} \ast A_{2})+((1-\alpha_{1}-\alpha_{2}) \ast (D_{v}-A_{1}-A_{2}))}{|D_{v}|}
    \end{split}
    \end{aligned}
    \end{equation}
where $|D_{v}|$ is the number of the nearest neighbors of node $v$, $\alpha_{1}$ and $\alpha_{2}$ are two tunable parameters in range (0,1), $A_{1}$ represents the number of neighboring nodes which $neighbor-diversity$ values are greater than node $v$, $A_{2}$ represents the number of neighboring nodes which $neighbor-diversity$ values are equal to node $v$.

The improved H-index centrality is proposed based on the dispersion degree of nodes, and the grade of neighboring nodes is reclassified according to the
relationship between neighboring nodes and the initial node. The nature of all neighboring nodes is considered comprehensively, and the improved method avoids the problem that nodes with different importance have the same level.

\begin{defn}
{\em
Cumulative function vector of node $v$ ($S(v)$), referring to the cumulative value of the neighbors of node $v$ at different $IH$ values in reversing order, is defined as:
}
\end{defn}
    \begin{equation}\label{adjmatix}
    \begin{aligned}
    \begin{split}
    S(v)=\{ c_{1}(v), c_{2}(v), c_{3}(v), ..., c_{h}(v) \}
    \end{split}
    \end{aligned}
    \end{equation}
where $c_{1}(v)$ is the largest $IH$ value of node $v$ neighborhood, $c_{h}(v)$ is the smallest $IH$ value of node $v$ neighborhood.
The node $v$ cumulative centrality is expressed as:
    \begin{equation}\label{adjmatix}
    \begin{aligned}
    \begin{split}
    MC(v)=\displaystyle \sum_{j=1}^{|N_{v}|}s^{1+j \ast \frac{j}{r}} S_{j}(v)
    \end{split}
    \end{aligned}
    \end{equation}
in this equation, $N_{v}$ is the set of neighbors of node $v$ and $|N_{v}|$ is the length of $N_{v}$, $s$ and $r$ are two tunable parameters in range $(0,1)$. $S_{j}(v)$ is the location in the sequence $S(v)$.
    \begin{equation}\label{adjmatix}
    \begin{aligned}
    \begin{split}
    IMH(v)=\displaystyle \sum_{j \in N_{v}} MC(v_{j})
    \end{split}
    \end{aligned}
    \end{equation}
where $N_{v}$ is the set of neighbors of node $v$.

Reversing ranking list according to the $IH$ value of the neighboring nodes of node $v$, since $IH$ value gap between nodes is not very great, and the high-order
cumulative value difference of different nodes, we use the parameter $s^{1+j \times \frac{j}{r}}$ in Eq.(3.4) to adjust the influence of each node, this
ensures that nodes with different $IH$ values also have the same important role in the specification.

In the next step, the extended mixing H-index centrality of node $v$ is on the basis of cumulative centrality. Eq.(3.6) is used for this goal:
    \begin{equation}\label{adjmatix}
    \begin{aligned}
    \begin{split}
    EMH(v)=IMH(v)+\displaystyle \sum_{j \in N_{v}}IMH(v_{j})
    \end{split}
    \end{aligned}
    \end{equation}
where $N_{v}$ is the set of nearest neighbors of node $v$.

Algorithm gives the concrete steps taken for implementation of the proposed  method, which is used to calculate the spreading capability of nodes and rank them. In this paper, $\alpha_{1}=0.5$, $\alpha_{2}=0.3$, $s=0.5$, $r=10$.

\begin{table}
\begin{center}
\begin{tabular}{l}
 \toprule
 Algorithm: Ranking nodes on the basis of cumulative centrality\\
 \midrule
 $\mathbf{01}$~~~~~~~~~~~~~~~~$\mathbf{Input}$:~~~~$G=(V,E)$~~~//$|V|=n$ and $|E|=N$\\
 $\mathbf{02}$~~~~~~~~~~~~~~~~$\mathbf{Output}$:~~Influential nodes ranking list\\
 $\mathbf{03}$~~~~~~~~~~~~~~~~$\mathbf{Begin~~Algorithm}$\\
 $\mathbf{04}$~~~~~~~~~~~~~~~~~~~~~~$\mathbf{for}$~~i=1~~$\mathbf{to}$~~n~~$\mathbf{do}$\\
 $\mathbf{05}$~~~~~~~~~~~~~~~~~~~~~~~~~~~~calculate $s_{j}(v_{i})$ using $Eq.(3.3)$\\
 $\mathbf{06}$~~~~~~~~~~~~~~~~~~~~~~$\mathbf{end}$~~$\mathbf{for}$\\
 $\mathbf{07}$~~~~~~~~~~~~~~~~~~~~~~$\mathbf{for}$~~i=1~~$\mathbf{to}$~~n~~$\mathbf{do}$~~~~~~~~~\\
 $\mathbf{08}$~~~~~~~~~~~~~~~~~~~~~~~~~~~~$MC(v_{i}) \leftarrow$ 0\\
 $\mathbf{09}$~~~~~~~~~~~~~~~~~~~~~~~~~~~~$MC(v_{i}) \leftarrow MC(v_{i})+s^{1+j \ast \frac{j}{r}} \cdot s_{j}(v_{i})$\\
 $\mathbf{10}$~~~~~~~~~~~~~~~~~~~~~~$\mathbf{end}$~~$\mathbf{for}$\\
 $\mathbf{11}$~~~~~~~~~~~~~~~~~~~~~~$\mathbf{for}$~~i=1~~$\mathbf{to}$~~n~~$\mathbf{do}$\\
 $\mathbf{12}$~~~~~~~~~~~~~~~~~~~~~~~~~~~~calculate $IMH(v_{i})$ using $Eq.(3.5)$\\
 $\mathbf{13}$~~~~~~~~~~~~~~~~~~~~~~$\mathbf{end}$~~$\mathbf{for}$\\
 $\mathbf{14}$~~~~~~~~~~~~~~~~~~~~~~$\mathbf{for}$~~i=1~~$\mathbf{to}$~~n~~$\mathbf{do}$\\
 $\mathbf{15}$~~~~~~~~~~~~~~~~~~~~~~~~~~~~$EMH(v_{i}) \leftarrow IMH(v_{i})$\\
 $\mathbf{16}$~~~~~~~~~~~~~~~~~~~~~~~~~~~~$\mathbf{foreach}$ $v_{j} \in N_{i}$ $\mathbf{do}$\\
 $\mathbf{17}$~~~~~~~~~~~~~~~~~~~~~~~~~~~~~~~~~~$EMH(v_{i}) \leftarrow EMH(v_{i})+IMH(v_{j})$\\
 $\mathbf{18}$~~~~~~~~~~~~~~~~~~~~~~~~~~~~$\mathbf{end}$~~$\mathbf{for}$\\
 $\mathbf{19}$~~~~~~~~~~~~~~~~~~~~~~$\mathbf{end}$~~$\mathbf{for}$\\
 $\mathbf{20}$~~~~~~~~~~~~~~~~$\mathbf{End~~Algorithm}$\\
 \bottomrule
\end{tabular}
\end{center}
\end{table}

In the algorithm, using the neighbor's dispersion of only one node cannot distinguish their influence values. At the same time, due to the low diversity and distribution of their nodes in the network, the influence of each node has a small diffusion range. Therefore, we consider the dispersion of all neighbor nodes. On account of the H-index centrality only considers the number of neighbors with high-quality information, minor response in a few neighbors do not cause changes in the overall spreading capacity of nodes. For other centrality measures, such as degree centrality and betweenness centrality, a few missing edges will have a significant impact on the ranking results. However, H-index centrality always assigns the same value to nodes of varying importance, leading to resolution constraints in distinguishing the actual spreading capability of these nodes. Aiming at the deficiency of H-index, the improved H-index centrality is proposed. Meanwhile, the cumulative centrality is also used in the algorithm, and the performance of neighbor nodes is fully considered. Different from H-index, this measure is based on all neighbor node's information to determine the spreading capability of the node, thus improving the accuracy and correctness of this method compared with other methods.

%================================L=a=p=l=a=c=i=a=n======================================================================================================================
\section{Analysis of network datasets and experimental results}
\subsection{Assessment Strategy}
One way to evaluate the ability of identifying and ranking nodes in networks is to distinguish the spreading capability of different nodes and distribute nodes uniformly at different levels. It has become a mainstream trend to use monotonicity to judge the capability of the proposed method and existing measures in the importance of different nodes \cite{20,29,30,31}. Eq.(4.1) is used to calculate the value of monotonicity. In this equation, $I$ is the ranking sequence of the measure, $N$ is the number of nodes in the network, $i$ is an element of list $I$, $N_{i}$ is the number of nodes in the same level of ranking list $N$. The range of $M(I)$ between $0$ and $1$, the larger the value, the better discriminating capability of the measure will be.
    \begin{equation}\label{adjmatix}
    \begin{aligned}
    \begin{split}
    M(I)=\left (1-\frac{\sum_{i \in I} N_{i} \times (N_{i}-1)}{N \times (N-1)}\right)^{2}
    \end{split}
    \end{aligned}
    \end{equation}

The accuracy of the proposed method is also an important evaluation standard \cite{32,33,34}. Susceptible-Infected-Recovered ($SIR$) spreading model is an abstract model widely used, and the concept of real dataset is also first proposed. In the $SIR$ spreading model, every node in the network can be divided into three different states: Susceptible ($S$), Infected ($I$), and Recovered ($R$). At the initial stage, only one node will be set to the infected state while all other nodes will be set to the susceptible state. Then, the initial infected node will propagate with probability $\alpha$ to all the other nearest neighboring nodes, while the infected nodes will recover with probability $\beta$ and be marked as the recovery state. In the end of the spreading process, there are only two states in the whole network: susceptible state and recovered state.

After estimating the spreading capability of each node, we use Kendall coefficient $\tau$ to express the relationship between the measure and the truthful data \cite{35,36}. If the value is larger, it means that the ranking list is closer to the actual value. The ranking list $M$ represents the ranking sequence of centrality measure, and $N$ represents spreading capability of nodes in the network, let ($m_{i}, n_{i}$) be a set of ordered pairs from two ranking sequences $M$ and $N$ respectively. For any set of sequences $(m_{i}, n_{i})$ and $(m_{j}, n_{j})$, if both $m_{i} > m_{j}$ and $n_{i} > n_{j}$ or if both $m_{i} < m_{j}$ and $n_{i} < n_{j}$, they are called to be concordant. If $m_{i} > m_{j}$ and $n_{i} < n_{j}$ or if $m_{i} < m_{j}$ and $n_{i} > n_{j}$, they are called to be discordant. If $m_{i} = m_{j}$ or $n_{i} = n_{j}$, they are not taken into account. The definition of kendall's coefficient $\tau$ is as follows:
    \begin{equation}\label{adjmatix}
    \begin{aligned}
    \begin{split}
    \tau = \frac{2(R_{a}-R_{b})}{R(R-1)}
    \end{split}
    \end{aligned}
    \end{equation}
where, $R_{a}$ denotes the number of concordant pairs and $R_{b}$ denotes the number of discordant pairs, $R$ denotes the number of all pairs.

In order to more clearly show the consequence of various centrality measures in different networks, we use Eq.(4.3) to calculate the mean value of all the conditions larger than the threshold $\beta_{th}$ to obtain a more convincing result \cite{20,37}.
    \begin{equation}\label{adjmatix}
    \begin{aligned}
    \begin{split}
    \sigma_{(\tau_{C})}=\frac{1}{N} \displaystyle \sum_{\beta_{min}=\beta_{th} + \delta}^ {\beta_{max}=\beta_{th} +\delta \times N} \tau_{C}
    \end{split}
    \end{aligned}
    \end{equation}
where $\tau_{C}$ represents the kendall coefficient of the centrality measure, $\beta_{th}$ denotes the epidemic threshold of network, $N = 10$ incremental step of $\beta$ value, $\delta = 0.01$.

Another measure used to test the performance of the proposed measure is to compare it with existing methods and calculate its improvement percentage ($\eta(\%)$) \cite{25,38}. The value obtained by Eq.(4.4) can accurately reflect the distinction between the two methods, and its definition is as follows:
    \begin{equation}\label{adjmatix}
    \begin{aligned}
    \begin{split}
    \eta(\%)=
    \left\{
    \begin{aligned}
    \frac{\tau_{\sigma(I)}-\tau_{I}}{\tau_{I}} \times 100, \tau_{I}>0\\
    \frac{\tau_{\sigma(I)}-\tau_{I}}{\tau_{I}} \times 100, \tau_{I}<0\\
    0, \tau_{I}=0\\
    \end{aligned}
    \right.
    \end{split}
    \end{aligned}
    \end{equation}
where $\tau_{\sigma(I)}$ denotes the kendall $\tau$ value of the measure $EMH$, $\tau_{I}$ denotes the kendall $\tau$ value of the other measures. When $\eta(\%) > 0$, $EMH$ is better than existing methods; when $\eta(\%) < 0$, existing methods are better than $EMH$; and when $\eta(\%) = 0$, it means that $EMH$ has no obvious difference compared with existing methods.

\subsection{Network datasets}
\begin{table}
Table 1: statistic properties of the used network datasets.

\begin{tabular}{cccccccc}
 \toprule
 Network & |V| & |E| & Average number & Maximum degree & $\alpha$ & $\mu$ & Assortativity\\
 \midrule
 Dolphins & 62 & 159 & 5.129 & 12 & 0.9 & 0.2 & -0.0436\\
 Polbooks & 105 & 441 & 8.400 & 25 & 0.8 & 0.4 & -0.1279\\
 Jazz & 198 & 2742 & 27.967 & 100 & 0.9 & 0.2 & 0.0202\\
 USair & 332 & 2126 & 12.81 & 139 & 0.9 & 0.2 & -0.2079\\
 Email & 1133 & 5451 & 9.622 & 71 & 0.9 & 0.2 & 0.0782\\
 WS & 2000 & 6012 & 6.021 & 11 & 0.6 & 0.6 & -0.0563\\
 LFR-2000 & 2000 & 4997 & 9.988 & 39 & 0.2 & 0.9 & -0.0032\\
 Yeast & 2361 & 7181 & 6.083 & 65 & 0.9 & 0.2 & -0.0489\\
 \bottomrule
\end{tabular}
\end{table}

In this paper, a total of six real network datasets and two artificial network datasets were involved. Among them, the real network datasets including Lusseau's Bottlenose Dolphins social network (Dolphins) \cite{39}, the network of selling political books about the presidential election in Amazon during 2004 (Polbooks) \cite{40}, the network of different Jazz musicians relationships (Jazz) \cite{41}, USair transportation network (USair) \cite{42}, the network of exchanging e-mail messages between members in the University Rovira Virgili (Email) \cite{43}, a social network which represents protein-protein interaction (Yeast) \cite{44}. In artificial network datasets, including Small-World network (WS) \cite{45} and Lancichinetti-Fortunato-Radicchi network (LFR) \cite{46}, both sets of artificial network datasets are generated by software Gephi. The specific data of the network used in this paper are shown in Table 1.

\subsection{Researching for the performance of $EMH$ from the perspective of spreading dynamics}
This experiment is used to examine the performance of various centrality measures. In order to find out the most important spreaders in the network, the spreading capability of nodes in various measures need to be calculated. Statistic centrality measures, SIR epidemic model and kendall $\tau$ ranking correlation coefficient have been used to execute this experiment. We use six real networks and two artificial networks as shown in Table 1 to conduct experiments, and compare the other seven existing centrality measures involved in section 2. The results of the comparison experiment are shown in Fig. 1 and Fig. 2. In these figures, X-axis represents the different infection rates in the SIR spreading model, Y-axis represents the kendall $\tau$ ranking relationship between various centrality measures and spreading capability of nodes, and the dotted line represents the threshold $\beta_{th}$ value of different networks.

The experimental result in Table 2 clearly shows the performance of various centrality measures under different experimental network datasets, and the results obtained by calculating the average kendall coefficient $\tau$ under the condition that the infection rate $\beta$ is greater than the threshold $\beta_{th}$ are more convincing. Table 2 shows that the average $\tau$ number $\tau_{\sigma(I)}$ of the proposed measure $EMH$ is greater than the existing centrality measures except in USair network, it means that $EMH$ is superior to existing measures in local performance.

    \begin{figure}
        \includegraphics[width=17cm]{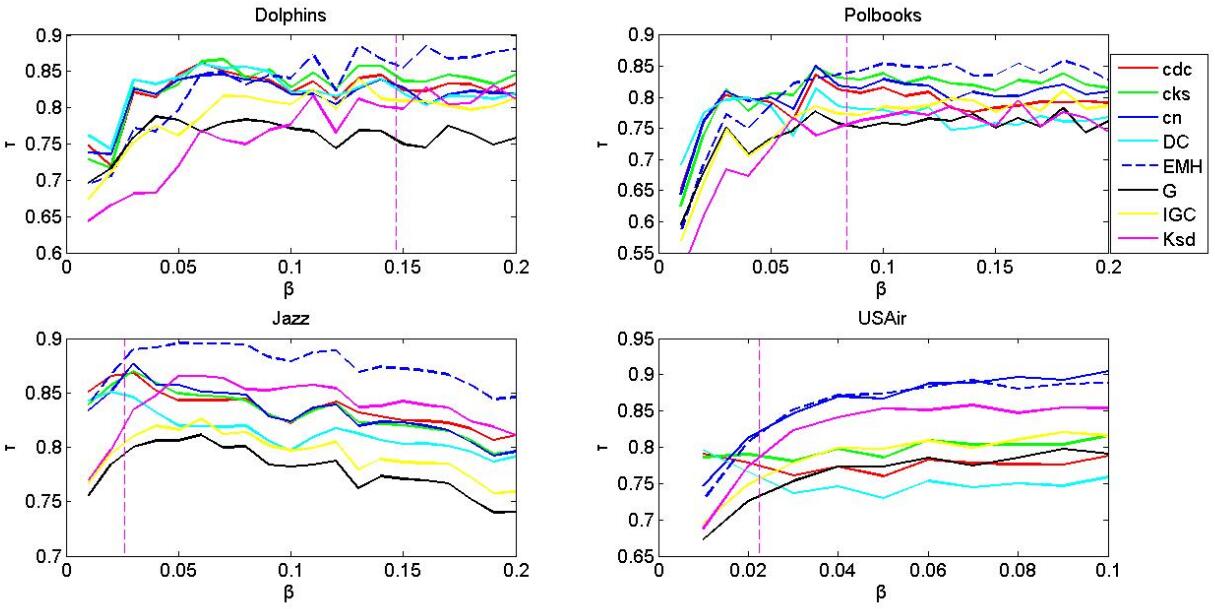}\\
        \caption{Kendall $\tau$ value curve under different infection and recovery rates on Dolphins, Polbooks, Jazz, USair networks.}\label{1}
    \end{figure}
    \begin{figure}
        \includegraphics[width=17cm]{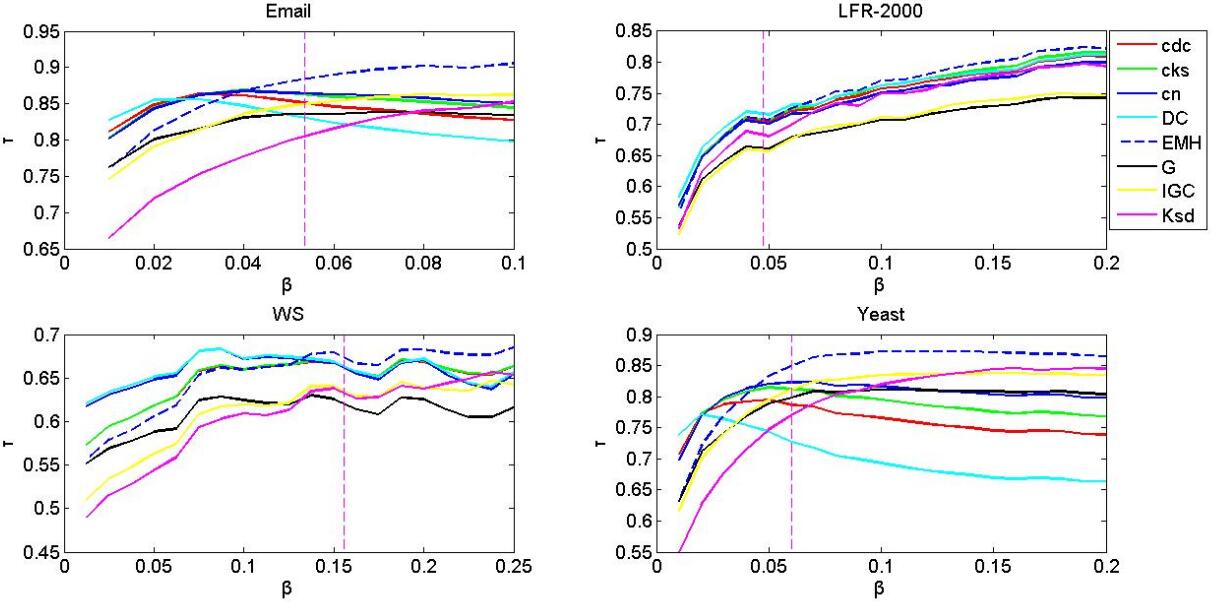}\\
        \caption{Kendall $\tau$ value curve under different infection and recovery rates on Email, WS, LFR-2000, Yeast networks.}\label{1}
    \end{figure}

In order to further investigate the relationship between the accuracy of $EMH$ and different $\beta$ values, it has been varied around $\beta_{th}$ were selected for verification in the experiment. By comparing the kendall $\tau$ value curve between the proposed method $EMH$ and the existing measures, the global effect between the involved methods could be more clearly reflected. Kendall $\tau$ value curve over the network datasets are shown in Fig. 1 and Fig. 2. The experimental results demonstrate that $cn$, $cks$ and $cdc$ methods are closer to the proposed method $EMH$ than other methods at the very start. With the increasing of $\beta$ and exceeding $\beta_{th}$, $EMH$ exhibits greater accuracy than in the other existing measures.

Hence, from this experiment, we can conclude that the proposed method $EMH$ is superior to existing methods in judging node's spreading efficiency and has good practicability.

\begin{table}
Table 2: the average kendall coefficient $\tau$ value ($\sigma(\tau_{(I)})$) for the networks.

\small
\begin{tabular}{cccccccccc}
 \toprule
 Network&$\sigma(\tau_{cdc})$&$\sigma(\tau_{cks})$&$\sigma(\tau_{cn})$&$\sigma(\tau_{DC})$&$\sigma(\tau_{EMH})$&$\sigma(\tau_{G})$&$\sigma(\tau_{IGC})$&$\sigma(\tau_{Ksd})$\\
 \midrule
 Dolphins & 0.827835 & 0.839498 & 0.818724 & 0.814254 & $\mathbf{0.872515}$ & 0.756770 & 0.805420 & 0.732248\\
 Polbooks & 0.793113 & 0.824586 & 0.811343 & 0.764427 & $\mathbf{0.845477}$ & 0.761273 & 0.788228 & 0.720380\\
 Jazz & 0.832814 & 0.829800 & 0.831224 & 0.810278 & $\mathbf{0.877774}$ & 0.779796 & 0.795255 & 0.743151\\
 USair & 0.774440 & 0.800168 & $\mathbf{0.881734}$ & 0.745820 & 0.878680 & 0.779242 & 0.803737 & 0.772300\\
 Email & 0.836538 & 0.852249 & 0.857269 & 0.809721 & $\mathbf{0.898734}$ & 0.836388 & 0.859894 & 0.815252\\
 WS & 0.661434 & 0.662105 & 0.656615 & 0.658641 & $\mathbf{0.680040}$ & 0.617052 & 0.638210 & 0.120744\\
 LFR-2000 & 0.767060 & 0.772502 & 0.759060 & 0.771385 & $\mathbf{0.779905}$ & 0.713946 & 0.718948 & 0.417018\\
 Yeast & 0.756931 & 0.786108 & 0.809755 & 0.684121 & $\mathbf{0.869484}$ & 0.807482 & 0.832720 & 0.721524\\
 \bottomrule
\end{tabular}
\end{table}

\subsection{Investigation of performance of $EMH$ in terms of improvement percentage}
This experiment is used to evaluate the proportion of improvement between the proposed method and existing measures. The larger the value calculated by
Eq.(4.4), the larger the gap between the comparison measures. Experimental results are shown in Fig. 3 and Fig. 4. The experimental results demonstrate that the $\eta(\%)$ value changes from negative to positive near the threshold $\beta_{th}$, and gradually improves with the increase of $\beta$ value. However, in USair network, the performance of $EMH$ is slightly inferior to the traditional $cn$ centrality due to the network characteristics. The above experimental results conclude that the proposed method $EMH$ is highly competitive in percentage improvement $\eta(\%)$.

\begin{figure}
        \includegraphics[width=17cm]{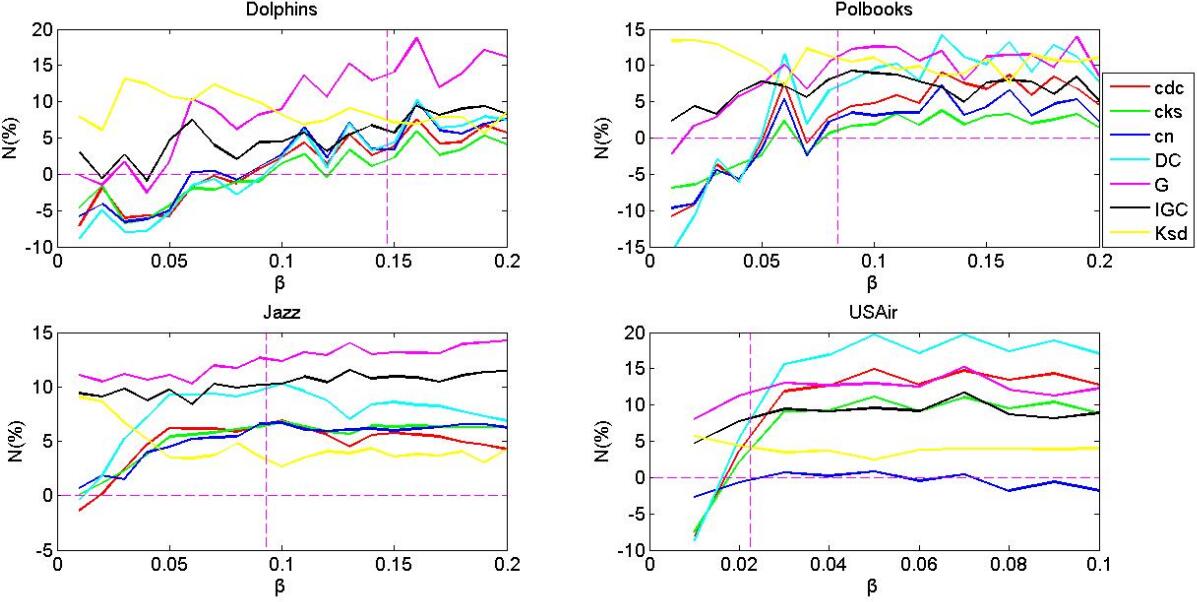}\\
        \caption{the improved $\tau$ percentage $\eta$(\%) of the proposed measure over the other indexing methods for Dolphins, Polbooks, Jazz, USair networks.}\label{1}
    \end{figure}
    \begin{figure}
        \includegraphics[width=17cm]{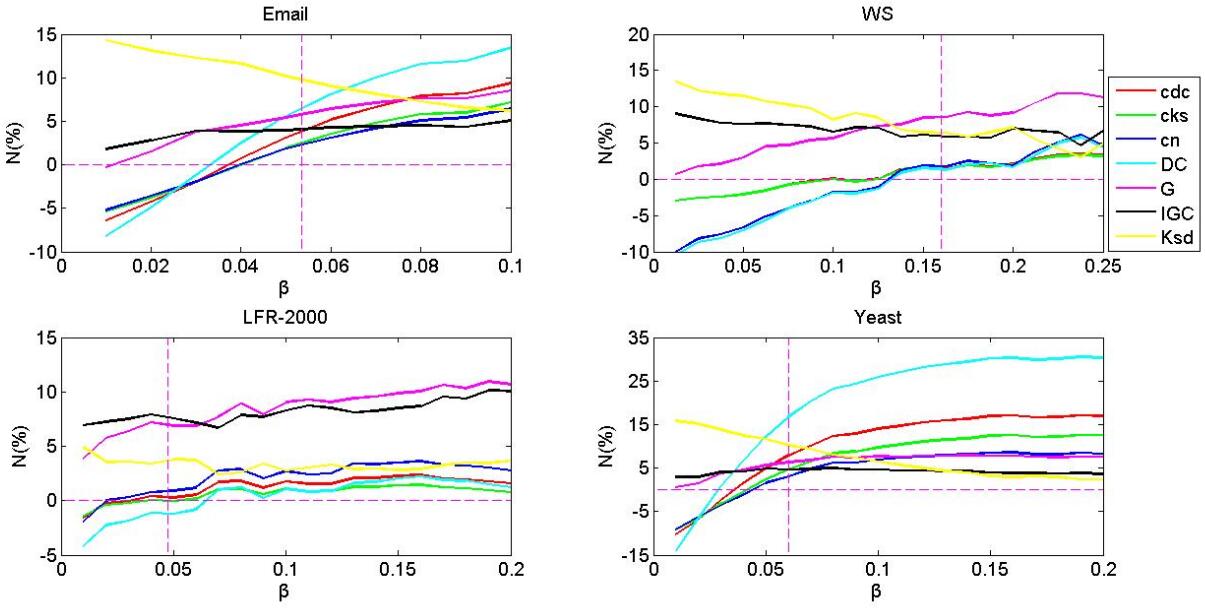}\\
        \caption{the improved $\tau$ percentage $\eta$(\%) of the proposed measure over the other indexing methods for Email, WS, LFR-2000, Yeast networks.}\label{1}
    \end{figure}

\subsection{Investigation of ranking monotonicity by centrality measures}
This experiment investigates the performance of the proposed method from a special perspective, that is, how many different levels of nodes in the network can be divided by centrality measures. This research method is proposed because the existing measures have the same level of nodes with different spreading capability when identifying the importance of nodes. The experimental results are presented in Table 3. Existing measures $cdc$, $cks$, $cn$ and $ksd$ have the same well performance as the proposed method $EMH$ in part of networks. On the whole, $EMH$ is better than the existing methods in distinguishing the importance of nodes, and can divide nodes in the whole network more widely.

\begin{table}
Table 3: The M value of ranking list generated by different measures in ten networks.

\begin{tabular}{ccccccccc}
 \toprule
 Network & M(cdc) & M(cks) & M(cn) & M(DC) & M(EMH) & M(G) & M(IGC) & M(Ksd)\\
 \midrule
 Dolphins & 0.9905 & 0.9905 & 0.9284 & 0.8312 & $\mathbf{0.9979}$ & 0.9916  & 0.9947 & $\mathbf{0.9979}$\\
 Polbooks & 0.9998 & $\mathbf{0.9999}$ & 0.9641 & 0.8252 & $\mathbf{0.9999}$ & 0.9982  & 0.9993 & $\mathbf{0.9999}$\\
 Jazz & 0.9994 & 0.9994 & 0.9982 & 0.9659 & $\mathbf{0.9998}$ & 0.9995  & 0.9995 & 0.9995\\
 USair & 0.9942 & 0.9942 & 0.9628 & 0.8586 & $\mathbf{0.9957}$ & 0.9942  & 0.9949 & 0.9951\\
 Email & 0.9990 & 0.9990 & 0.9839 & 0.8875 & $\mathbf{0.9999}$ & 0.9996  & 0.9998 & $\mathbf{0.9999}$\\
 WS & 0.9959 & 0.9957 & 0.6085 & 0.5922 & $\mathbf{0.9999}$ & 0.9757  & 0.9982 & 0.9998\\
 LFR-2000 & $\mathbf{0.9999}$ & $\mathbf{0.9999}$ & 0.9789 & 0.8760 & $\mathbf{0.9999}$ & 0.9997  & 0.9998 & $\mathbf{0.9999}$\\
 Yeast & 0.9921 & 0.9921 & 0.9458 & 0.7210 & $\mathbf{0.9994}$ & 0.9959  & 0.9964 & 0.9965\\
 \bottomrule
\end{tabular}
\end{table}

%==============================s=i=n=g=l=e=s=s====L=a=p=l=a=c=i=a=n===================================================================================================================================================================================================================

\section{Summary and Discussion}
In recent years, with the ever-increasing scale of users and using frequency of social network have made this measure become the best choice of news dissemination and exchange of information. The primary mission in the use of these social networks is to rank nodes and find out the most valuable spreaders. In this paper, a centrality measure is defined based on the dispersion degree, diffusion range, influence intensity and cumulative centrality of neighboring nodes, and the proposed method is applied to rank the importance of nodes by neighborhood topological structure. The experimental results on a series of networks indicate that $EMH$ is superior to the existing methods in distinguishing the importance and influence of nodes. The proposed method has high accuracy and availability with great research value and significance.

\medskip

%\noindent \textbf{Acknowledgements}~~The authors would like to thank Dr. Xiaogang Liu for his suggestion of constructing infinite families of $A$-integral graphs by using the subdivision-vertex and subdivision-edge coronae.

\end{document}